\documentstyle[11pt,newpaspmy,epsf]{article}
\begin{document}
\title{New  radio observations of QSO 0957+561 confirm optical-to-radio time 
delay}
\author{V. L. Oknyanskij}
\affil{ Sternberg State Astronomical Institute,
        Moscow State University, Universitetskij Prospekt 13,
        Moscow, 119899, Russia; E-mail: oknyan@sai.msu.ru}

\begin{abstract}

In our previous publications have been reported about possible time delay between optical and 
radio (6 cm) variations in QSO 0957+561 and noted that the result can be tested with new radio 
observations. Here we have made this test using new published (Haarasma et al., 1999) radio 
observations of the object in 6 cm and 4 cm.  We have found that the new observations confirm 
optical-to-radio time delay. Additionally we have found that radio 6 cm variations followed 4 cm 
ones with time delay about 230 days.

Obtained results imply  existence of the variable beamed hard radiation in the nucleus of the 
QSO.

\end{abstract}

\section{Introduction}

\begin{figure}
\vspace{2cm}
\plotfiddle{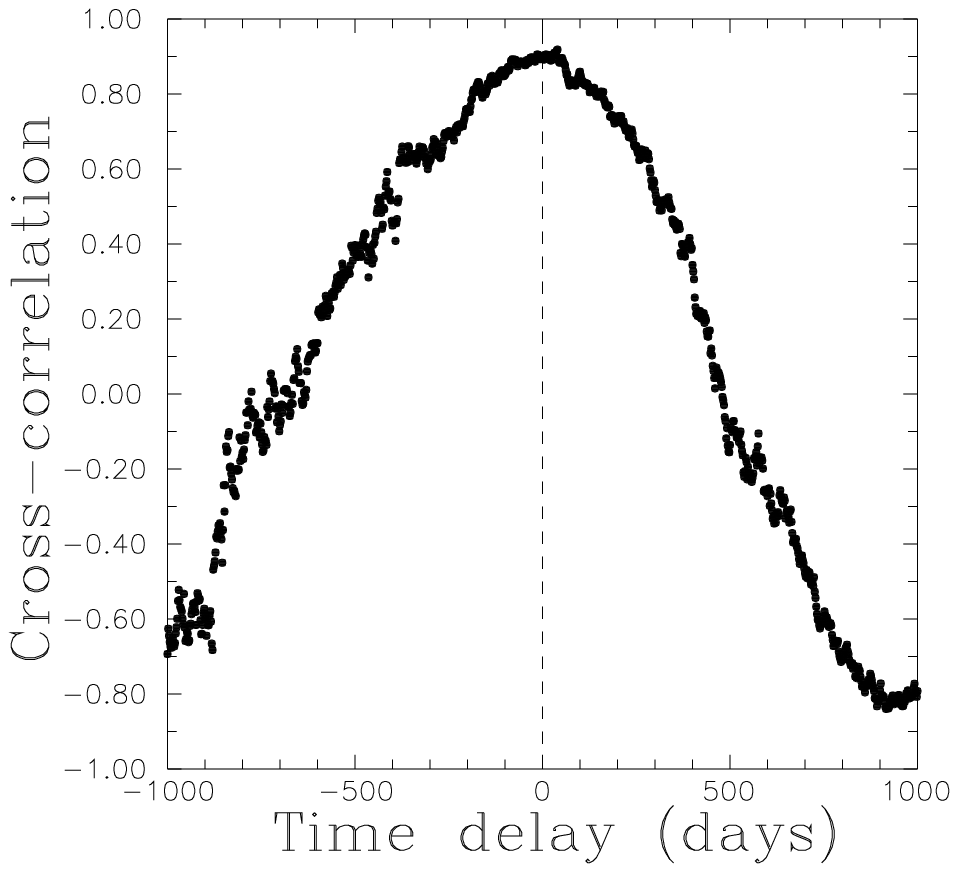} {5 cm}{0}{70}{70}{-230}{-185}
\caption{Cross-correlation between radio 4 cm (shifted ahead by 230 days) and 
6 cm combined light curves. The figure shows that 4 cm and 6 cm data
correlate very well after this shifting and can be combined in one light
curve}
\end{figure}

\begin{figure}
\vspace{2cm}
\plotfiddle{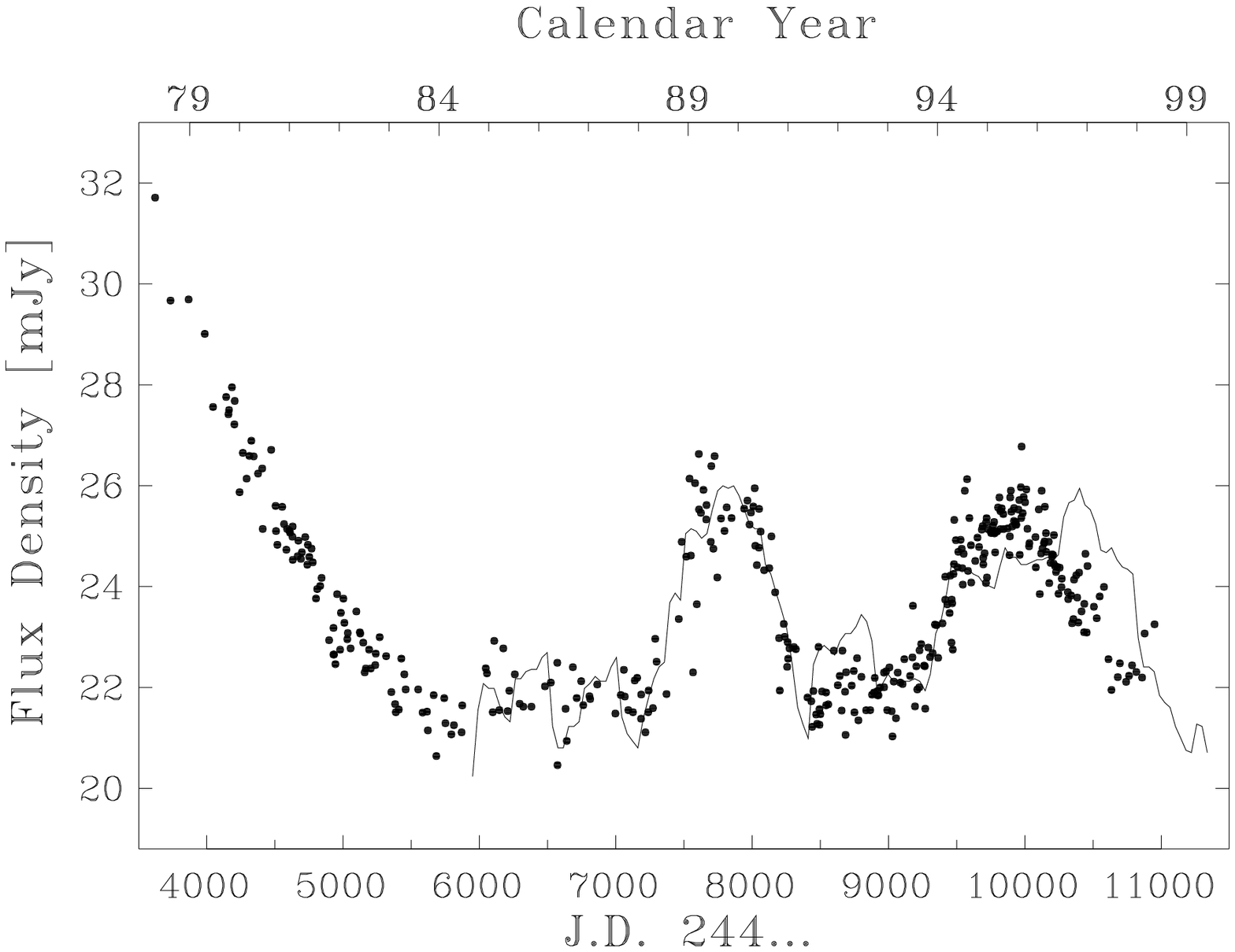} {5 cm}{0}{70}{70}{-185}{-185}
\caption{Radio (points)  and optical (line) combined light curves. 
The optical light curve  have been transferred to the radio light curve system  
(see details in Oknyanskij, 1997)  by (i) smoothing in 200 days intervals
(ii) taking into account that radio variations have time delay to optical ones 
which is a linear function of time, and that (iii) radio flux  response  is a 
power-low function of time with power  
coefficient $a= -0.62$. The radio combine light curve include 6 cm and 4 cm 
data. The 4 cm radio data have been shifted ahead by 230 days.}
\end{figure}

Radio-optical variability correlation in QSO 0957+561  (first gravitationally lensed system), was 
first reported by Oknyanskij   and Beskin (1993), on  the basis of VLA  (6 cm) radio 
observations made in the years 1979 to 1990. 
Then we have found that the time  delay  value  was a linear function of time 
with the rate of increase being  about 110  days/year  while the portion 
of reradiate flux in the radioresponse was  some decreasing function of time 
(Oknyanskij, 1997,1998). Our  results allowed predictions to be made for radio
variations for the years  1995  and latter.

\section{Results}

At the present time, new VLA radio observations (at 4 cm and 6 cm) of QSO  0957+561 during 
1995-1997 have been published (Haarsma et al.  1999) by  which we may check the predictions. 
 We have found that 6 cm variations followed 4 cm ones with delay about 230 days (see Fig.1). 
We take this fact into account as well as gravitational lensing time delay between variations in A 
and B  to have got combined radio light curve.  As we have found the radio light curve during 
these 3 years have been following our predictions in limits of the estimated standard 
errors (see Fig.2). 

\section{Conclusion}

Obtained results indicate that the optical and radio emitting regions are physically related, but 
have distinct size scales, locations and possibly radiation   mechanisms.
We  conclude that the variable radio source is ejected from the central  part of the QSO  compact 
component (Oknyanskij, 1999). Obtained results imply  existence of the variable beamed hard radiation in the nucleus of the 
QSO.

\subsection*{Acknowledgements}
Financial support from the Russian Foundation for Basic Research
through grant No 01-02-16800 is gratefully acknowledged.


\begin{references}
  
  \reference  Haarsma, D.B., et al. 1999, \apj,  510, 64
  \reference  Oknyanskij, V.L., Beskin, G.M. 1993, in
    {\it Proceedings of the
  31th Liege International Astriphys. Call.}, p.65
  \reference  Oknyanskij, V.L. 1997, \apss, 246, 299
  \reference  Oknyanskij, V.L. 1998, \apss, 259, 295
  \reference Oknyanskij, V.L. 1999,  Astron. Nachr., 320, 314

\end{references}
\end{document}